\documentclass[aps,prd,superscriptaddress,onecolumn,amssymb,eqsecnum,showpacs,nofootinbib,showkeys]{revtex4-1}
\pdfoutput=1
\usepackage{bm}
\usepackage{amsmath}
\usepackage[latin1]{inputenc}
\usepackage[english]{babel}
\usepackage[T1]{fontenc}
\usepackage{amssymb}
\usepackage{amsfonts}
\usepackage{epsfig}
\usepackage{colordvi}
\usepackage{psfrag}
\usepackage{color}
\newcommand{\beq}{\begin{equation}}
\newcommand{\eneq}{\end{equation}}

\begin{document}

\title{String duality transformations in $f(R)$ gravity from Noether symmetry approach}

\author{Salvatore Capozziello}
\email{capozziello@na.inf.it} 
\affiliation{Dipartimento di Fisica, Università di Napoli ``Federico II'', Compl. Univ. di Monte S. Angelo, Edificio G, Via Cinthia, I-80126, Napoli, Italy} 
\affiliation{Istituto Nazionale di Fisica Nucleare (INFN), Sezione  di Napoli, Compl. Univ. di Monte S. Angelo, Edificio G, Via Cinthia, I-80126, Napoli, Italy}
\affiliation{Gran Sasso Science Institute (INFN), Via F. Crispi 7, I-67100, L'Aquila, Italy}

\author{Gabriele Gionti S. J.}
\email{ggionti@as.arizona.edu}
\affiliation{Specola Vaticana Vatican City, V-00120, Vatican City State and Vatican Observatory Research Group Steward Observatory, The University Of Arizona, 933 North Cherry Avenue Tucson, Arizona 85721, USA}
\affiliation{Istituto Nazionale di Fisica Nucleare (INFN), Laboratori Nazionali di Frascati, Via E. Fermi 40, 00044 Frascati, Italy}

\author{Daniele Vernieri}
\email{vernieri@iap.fr, {\bf corresponding author}}
\affiliation{Sorbonne Universit\'es, UPMC Univ Paris 6 et CNRS, UMR 7095, Institut d'Astrophysique de Paris, GReCO, 98bis Bd Arago, 75014 Paris, France}

\date{\today}
\begin{abstract}
We select $f(R)$ gravity models that undergo scale factor duality transformations. As a starting point, we consider the tree-level effective gravitational action of bosonic String Theory coupled with the dilaton field. This theory inherits the Busher's duality of its parent String Theory. Using conformal transformations of the metric tensor, it is possible to map the tree-level dilaton-graviton string effective action into $f(R)$ gravity, relating the dilaton field to the Ricci scalar curvature. 
Furthermore, the duality can be framed under the standard of Noether symmetries and exact cosmological solutions are derived. Using suitable changes of variables, the string-based $f(R)$ Lagrangians are shown in cases where the duality transformation becomes a parity inversion. 
\end{abstract}

\pacs{04.50.Kd, 98.80.-k, 11.25.Tq}
\keywords{Modified gravity, quantum cosmology, string theory and cosmology}

\maketitle

\section{Introduction}
\label{one}

In the last thirty years, several shortcomings came out in General Relativity (GR), essentially related  to the ultraviolet and infrared behaviors of the theory. In other words, while the theory works very well at intermediate scales (Solar System, up to Galactic scales), it shows problems at quantum and cosmological scales. In the first case, the lack of a Quantum Gravity Theory means that GR cannot be dealt under the standard of the other fundamental theories. In the other case, cosmic dynamics is addressed if huge amounts of dark energy and dark matter are assumed as sources in the Einstein field equations. The shortcoming consists in the fact that such exotic ingredients have not been detected at fundamental level up to now, despite of the fact that their effects are dramatically present at astrophysical and cosmological scales.
 
A change of perspective in this  state of art consists to resort to alternative theories of gravity considered as semi-classical schemes capable of addressing, at least in part, the above mentioned problems and of retaining  the positive results of GR.
Among these alternatives, one of the most fruitful approach has been the one called {\it Extended Theories of Gravity} (ETGs) based on corrections and enlargements of GR~\cite{Capozziello:2007ec,Capozziello:2011et,Nojiri:2010wj,libroSalvFara}. 
The paradigm consists, essentially, in adding minimally and/or non-minimally coupled scalar fields or higher-order curvature invariants into the dynamics. The reason of these additional terms is that any quantum field theory formulated on curved space gives rise to this kind of corrections~\cite{birrell}.

Furthermore, any unification scheme, as Superstring, Supergravity or Grand Unified Theories (GUT), takes in account effective actions where non-minimal couplings or higher-order terms in the curvature invariants are present. Specifically, we have to consider terms in the Einstein-Hilbert Lagrangian coming from one-loop or higher-loop corrections in the quantum regime~\cite{birrell,Nojiri:2003vn,Vilkovisky:1992pb}. 
In this framework, the so called $f(R)$ gravity theories~\cite{Capozziello:2011et} are the first straightforward extension of GR as soon as the condition $f(R)\neq R$ is requested.

The main criticism to this approach is that ETGs can be somehow phenomenological and adapted to particular issues coming, e.g., from cosmology or astrophysics (see~\cite{Capozziello:2012ie} for a review). However, specific features can be selected in order to match these models with  some fundamental theory. In particular, $f(R)$ gravity models showing duality invariance could be useful to relate ETGs with Superstring Theory in the low-energy limit~\cite{Becker,Green,Veneziano:1991ek,Scherk:1974ca,Meissner:1991zj,Gasperini:1991qy,Gasperini:1992em,Tseytlin:1991xk,Zwiebach}. An important output of this approach could be the geometrical interpretation of the dilaton which is a specific feature of String Theory. In general, looking for fundamental symmetries for effective models is a sort of {\it reconstruction procedure} useful to recover fundamental theories from phenomenology. 

The aim of this paper is to show that string-dilaton gravity and $f(R)$ gravity are related by conformal transformations and that scale factor duality can be recovered for some $f(R)$ gravity models. This renders possible to select physical $f(R)$ models at fundamental level. 
Another important issue addressed in this paper is related to the fact that duality can be derived from Noether symmetries~\cite{Capozziello:1996bi,Capozziello:1993tr}. In this sense, duality could be considered in relation to conserved quantities emerging in physical theories.  

The outline of the paper is as follows. In Section \ref{two}, the main features of Busher's duality in bosonic String Theory are summarized.  The tree-level effective bosonic String Theory inherits the Busher's duality transformations of its parent String Theory and holds $O(n,n)$ symmetry~\cite{Meissner:1991ge}. 
Section \ref{three} is devoted to $f(R)$ gravity which is mapped into the tree-level dilaton-graviton string effective action, with the dilaton field and potential being related to the $f(R)$ functions. 
Scale factor duality in $f(R)$ cosmology is discussed in Section \ref{four}. In particular, by means of Noether symmetries we derive $f(R)$ models for which the invariance under the duality transformation $a\rightarrow 1/a$ holds for the cosmic scale factor $a(t)$ at the level of the Lagrangian. We show that cosmological solutions can be derived for such $f(R)$ models, where Noether symmetries allow the reduction and the exact integration of the dynamical system. In Section \ref{five}, we draw conclusions and sketch the perspectives. A summary of the Noether Symmetry Approach is reported in the Appendix \ref{six}, giving particular emphasis to $f(R)$ gravity models. 

\section{Duality in string theory and its cosmological consequences}
\label{two}

Let us start our considerations taking into account duality in String Theory. It is well known that certain classes of two dimensional sigma model exhibits a duality symmetry. It can be shown that two distinct actions, having two geometrically inequivalent target spaces, describe the same physics. In fact, let us consider a two-dimensional sigma model embedded in a $(n+1)$ Lorentzian Manifold $\{{\cal M},G\}$. The action is~\cite{Buscher:1987qj}: 

\begin{equation}
S=\frac{1} {4 \pi {\alpha}'} \int d^{2}\xi  \Big (\sqrt{h}h^{\mu\nu}G_{ab}(\partial_{\mu}X^{a})(\partial_{\nu}X^{b})+\epsilon^{\mu\nu}B_{ab}(\partial_{\mu}X^{a})(\partial_{\nu}X^{b})
+{\alpha}' {\sqrt{h}}R^{(2)}\phi(X^{a})\Big),
\label{totstring}
\end{equation}
where $\alpha'$ is the Regge slope parameter, $X^{a}$ are the coordinates on the target space, $h$ is the absolute value of the determinant of the two-dimensional metric $h_{\mu\nu}$ on the world-sheet, $\epsilon_{\mu\nu}$ is the two-dimensional Levi-Civita tensor density,
$B_{ab}$ is an antisymmetric field, $R^{(2)}$ is the Ricci scalar curvature on the world-sheet, $\phi(X^{a})$ is the dilaton field. Being ${\cal L}$ the Lagrangian in the big parentheses under the sign of integral of Eq. (\ref{totstring}), in order to define the dual Lagrangian ${\tilde {\cal L}}$, there must exist a vector field $X$ which is an isometry for ${\cal L}$~\cite{Rocek:1991ps,Alvarez:1994dn}, that is $L_{X} {\cal L}=0$, where $L_{X}$ is the Lie derivative of ${\cal L}$ along $X$. In the case ${\displaystyle X=\frac{\partial}{\partial X^{0}}}$, the dual action is: 
\begin{equation}
{\tilde S}=\frac{1} {4 \pi {\alpha}'} \int d^{2}\xi  \Big (\sqrt{h}h^{\mu\nu}{\tilde G}_{ab}(\partial_{\mu}{\tilde X}^{a})(\partial_{\nu}{\tilde X}^{b})+\epsilon^{\mu\nu}{\tilde B}_{ab}(\partial_{\mu}{\tilde X}^{a})(\partial_{\nu}{\tilde X}^{b})
+{\alpha}' {\sqrt{h}}R^{(2)}\phi(X^{a})\Big),
\label{duatotstring}
\end{equation}
where 
\begin{equation}
{\tilde G}_{00}=1/G_{00}\,, \,\,\,\, {\tilde G}_{0i}=B_{0i}/G_{00}\,, \,\,\,\, {\tilde G}_{ij}=G_{ij} -  (G_{0i}G_{0j}-B_{0i}B_{0j})/G_{00}\,, \,\,\,\, i,j=1,...,n\,,
\label{medual}
\end{equation}
and 
\begin{equation}
{\tilde B}_{0i}=-{\tilde B}_{i0}=G_{0i}/G_{00}\,, \,\,\,\, {\tilde B}_{ij}=B_{ij}+(G_{0i}B_{0j}-B_{0i}G_{0j})/G_{00}\,, \,\,\,\, i,j=1,...,n\,.
\label{macca}
\end{equation}
Eqs.~(\ref{medual}) and~(\ref{macca}) are the so called {\it Busher's duality relations} for this particular isometry~\cite{Buscher:1987qj, Buscher:1987sk}. The dual action is dynamically equivalent to the original one at classical level. At quantum level, the conformal invariance of the model in Eq. (\ref{totstring}), at one loop, implies the imposition of  the following conditions~\cite{Green}:

\begin{eqnarray}
{\frac{1}{{\alpha}'}}\left({\frac{n-25}{48{\pi}^{2}}}\right)+{\frac{1}{16{\pi}^{2}}}\Big(4(\nabla \phi)^{2}-4{\nabla}^{2}{\phi}-R+{\frac{1}{12}}H^{2}\Big)=0\,, \nonumber \\
R_{ab} +2 {\nabla}_{a}{\nabla}_{b}\phi-\frac{1}{4}H_{a}^{\phantom{a}cd}H_{bcd}=0, \,\,\,\,\,\, {\nabla}^{c}H_{cab}-2\left({\nabla}^{c}\phi\right)H_{cab}=0\,.
\label{simant}
\end{eqnarray}
where $\nabla$ is the covariant derivative on the target space, $H_{abc}=\partial_{a}B_{bc}+\partial_{b}B_{ca}+\partial_{c}B_{ab}$, and $H^2=H_{abc}H^{abc}$. The Eqs.~(\ref{simant}) can be derived by the Euler-Lagrange equations of the following (effective) action:

\begin{equation}
S^{n+1}=K\int d^{n+1} x \sqrt{-G}e^{-2\phi}\left (R+4{\nabla}_{a}\phi{\nabla}^{a}\phi - {\frac{1}{12}}H^{2} + \Lambda\right),
\label{effetto}
\end{equation} 
where $K$ is a coupling constant, $n$ is the number of the spatial dimensions, and the cosmological constant term is given by $\Lambda = \frac{25-n}{3 \alpha'}$, which is equal to $0$ if and only if $n=25$ (the critical dimension for bosonic String Theory). 

If the original model (\ref{totstring}) satisfies the one-loop conformal Eqs.~(\ref{simant}), then so the dualized model (\ref{duatotstring}), provided that the dilaton field shifts in the following way~\cite{Buscher:1987qj}:
\begin{equation}
\phi \rightarrow {\tilde \phi}=\phi -\frac{1}{2}\ln {(g_{00})}.
\label{conf}
\end{equation}
In case the Lagrangian ${\cal L}$ has $d$-Abelian independent spatial isometries, then the invariance group of the action (\ref{totstring}) and its dual (\ref{duatotstring}) is the group $O(d,d)$~\cite{Maharana:2013uvy,delaOssa:1992vc,Gasperini:2002bn,DeAngelis:2013wba}. It has been proved~\cite{Meissner:1991ge} that, in case the moduli fields $G_{ab}$ and $B_{ab}$ depend only on time $t$, the effective action (\ref{effetto}) is $O(n,n)$ invariant. In Ref.~\cite{Gasperini:2002bn}, it has been shown that if the following spatially flat, homogeneous and isotropic metric (in $n+1$ dimensions)
\begin{equation}
ds^{2}=dt^{2}-a^{2}(t)dx^{2}_{i}\,,
\label{metro}
\end{equation}
where $a(t)$ is the scale factor, is a solution of the field equations, then also the metric with $a(t) \rightarrow a^{-1}(\pm t)$ is a solution. The effective gravitational action (\ref{effetto}) exhibits the following duality symmetric transformations among the solutions of its Euler-Lagrange equations (which are nothing else but an application of the Busher's duality relations in the case of a symmetry made out by $d$-Abelian spatial vector fields ${\displaystyle \frac {\partial}{\partial x^{i}}}$)
\begin{equation}
a \rightarrow \tilde{a} = a^{-1}, \,\,\,\,\, \phi \rightarrow {\tilde \phi}=\phi - d \ln(a)\,.
\label{cosmdual}
\end{equation} 
This is the {\it long-short} duality correspondence of the scale factor $a$ for the string-dilaton cosmology. The duality relations (\ref{cosmdual}), between the solutions of the Euler-Lagrange equations derived from the action (\ref{effetto}), allow to construct the so called {\it Pre-Big Bang} cosmological models~\cite{Gasperini:2002bn}. 
  
Let us consider now the \textit{tree-level dilaton-graviton string effective action} 
\beq
\mathcal{S}=\int d^Dx\sqrt{-G}e^{-2\phi}\left[R+4\nabla_\mu\phi\nabla^\mu\phi+\Lambda\right],   \label{2.0}
\eneq
that can be achieved from the above results, in the low-energy limit, retaining only the scalar mode (the dilaton) and the tensor mode (the graviton)~\cite{Becker,Green, Damour:1994ya, Polch, Veneziano:1991ek,Scherk:1974ca, Gasperini:1992em,Tseytlin:1991xk, Gasperini:2002bn}. Here $D=n+1$ is the total number of dimensions, and $G$ indicates the determinant of the $D$-dimensional spacetime metric.
In this sense, the effective string action reduce to a scalar-tensor theory (Brans-Dicke-like)~\cite{Brans:1961sx}. 
 
Starting from the action (\ref{2.0}) we can calculate the field equations for the gravitational field by varying the action with respect to the metric tensor, and the field equations for the dilaton field by varying the action with respect to the dilaton. 
Since we are interested in cosmological solutions describing the observable universe, we shall consider the case where the space-time has $D=4$ dimensions. In this case, strings are not in critical dimensions and $\Lambda \neq0$. In $4D$, we indicate, as usual, the metric tensor with $g_{\mu\nu}$ and its determinant with $g$.

So, by varying the action in Eq.~(\ref{2.0}) with respect to the metric tensor and the dilaton field we get, respectively, the field equations
\beq
G_{\mu\nu}=\frac{1}{2}\Lambda g_{\mu\nu}+2g_{\mu\nu}\Box\phi-2g_{\mu\nu}\nabla_\mu\phi\nabla^\mu\phi-2\nabla_\mu\nabla_\nu\phi\,,     \label{2.2}
\eneq 
and
\beq
\Box\phi=\nabla_\mu\phi\nabla^\mu\phi-\frac{1}{4}\left(R+\Lambda\right).    \label {2.1}
\eneq
Taking the trace of Eq.~(\ref{2.2}) we get
\beq
\Box\phi=-\frac{R}{6}+\frac{4}{3}\nabla_\mu\phi\nabla^\mu\phi-\frac{\Lambda}{3},  \label{2.4}
\eneq    
and comparing the latter equation with Eq.~(\ref{2.1}) we obtain 
\beq
\Box\phi=-\frac{1}{2}R,               \label{2.5}
\eneq
and also 
\beq
\nabla_\mu\phi\nabla^\mu\phi=\frac{1}{4}(\Lambda-R).      \label{2.6}
\eneq
The latter equation will be useful in the next Section.

\section{$f(R)$ gravity as a string-dilaton model}
\label{three}

The action of $f(R)$ gravity in $4D$ is given by the action  
\beq
 S=\int d^4x\sqrt{-g}f(R)\,.
\eneq
Varying the latter with respect to $g_{\mu\nu}$, one gets the field equations 
\beq
f'(R)R_{\mu\nu}-\frac{1}{2} f(R)g_{\mu\nu}R = \nabla_\mu\nabla_\nu f'(R)-g_{\mu\nu}\Box f'(R),    \label{1.1}  
\eneq
which are fourth-order partial differential equations in the metric due to the terms on the right side of the equation above, and the prime indicates the derivative with respect to $R$. By a suitable manipulation, it can be rewritten in an Einstein-like form as 
\beq
G_{\mu\nu} = R_{\mu\nu}-\frac{1}{2}g_{\mu\nu}R = \frac{1}{f'(R)}\left\{\frac{1}{2}g_{\mu\nu}\left[f(R)-Rf'(R)\right] 
+\nabla_\mu\nabla_\nu f'(R)-g_{\mu\nu}\Box f'(R)\right\},
\eneq
where the contributions due to the higher-order terms can be reinterpreted as a sort of curvature stress-energy tensor given by the form of the $f(R)$ function~\cite{Capozziello:2011et}. 
Considering also the standard perfect-fluid matter contributions and using physical units $8\pi G=c=1$, we have 
\beq
G_{\mu\nu}=\frac{1}{f'(R)}\left\{\frac{1}{2}g_{\mu\nu}\left[f(R)-Rf'(R)\right]+ \nabla_\mu\nabla_\nu f'(R)-g_{\mu\nu}\Box f'(R)\right\}+ \frac{T_{\mu\nu}^m}{f'(R)}=T_{\mu\nu}^{curv}+\frac{T_{\mu\nu}^m}{f'(R)}\,,             \label{1.2} 
\eneq
where $T_{\mu\nu}^{curv}$ is implicitly defined and $T_{\mu\nu}^{m}$ is the stress-energy tensor of the matter contributions. Moreover, $T_{\mu\nu}^{curv}$ identically vanishes as soon as $f(R)=R$ and the standard minimal coupling with matter is automatically recovered.
The trace of the Eq.~(\ref{1.2}) is
\beq
R f'(R)-2f(R)+3\Box f'(R)=T^{m}, \label{trace}
\eneq 
where $T^{m}=g^{\mu\nu}T_{\mu\nu}^{m}$. Such an equation dynamically relates $R$ with $T^{m}$ while in the GR case the algebraic relation $R=-T^{m}$ holds~\cite{Capozziello:2007ec,libroSalvFara}.

The effective string-dilaton Lagrangian can be connected to the $f(R)$ gravity Lagrangian via the conformal transformation
\beq
g_{\mu\nu}(x)\rightarrow\tilde{g}_{\mu\nu}(x)=\Omega^2(x)g_{\mu\nu}(x).    \label{3.0}
\eneq
In fact the two actions can be mapped into each other as 
\begin{equation}
\sqrt{-g}e^{-2\phi}\left(R+4\nabla_\mu\phi\nabla^\mu\phi+\Lambda \right)=\sqrt{-\tilde{g}}f(\tilde{R})\,,
\end{equation}
which, by using Eq.~(\ref{3.0}), becomes
\begin{equation}
e^{-2\phi}\left(R+4\nabla_\mu\phi\nabla^\mu\phi+\Lambda \right)=\Omega^4f(\tilde{R}).                 \label{3.21} 
\end{equation}
Plugging  Eq.~(\ref{2.6}) into Eq.~(\ref{3.21}), and properly arranging the terms, the latter becomes
\beq
f(\tilde{R})=2\Lambda \Omega^{-4}e^{-2\phi}\,,
\eneq
which, by choosing
\beq
\Omega=e^{-\phi}\,,  \label{new}
\eneq
finally becomes
\beq
f(\tilde{R})=2\Lambda e^{2\phi}\,.     \label{3.23}
\eneq

\section{Scale factor duality in $f(R)$ cosmology}
\label{four}

Scale factor duality for the above models can be realized in cosmology assuming an homogeneous and isotropic universe, described by a Friedmann-Lema$\hat{\imath}$tre-Robertson-Walker (FLRW) metric, which in spherical coordinates is 
\beq
ds^2=dt^2-a^2(t)\left[\frac{dr^2}{1-kr^2}+r^2\left(d\theta^2 +\mbox{sin}^2 \theta d\phi^2\right)\right],
\eneq
where $a(t)$ is the scale factor and $k$ is the spatial curvature parameter. 
In this metric, the Ricci scalar reads 
\beq
R=-6\left[\frac{\ddot{a}}{a}+\left(\frac{\dot{a}}{a}\right)^2+\frac{k}{a^2}\right].            \label{4.14} 
\eneq
In order to derive the cosmological equations, one can define a canonical Lagrangian $\mathcal{L}=\mathcal{L}(a,\dot{a},R,\dot{R})$, where $\mathcal{Q}=\left\{a,R\right\}$ is the configuration space and $\mathcal{TQ}=\left\{a,\dot{a},R,\dot{R}\right\}$ is the related tangent bundle on which $\mathcal{L}$ is defined.
The corresponding point-like action is~\cite{Capozziello:2007ec,Capozziello:1996bi}
\beq
S=4\pi\int dt \mathcal{L}(a,\dot{a},R,\dot{R}),                                  
\eneq
where $a$ and $R$ are independent variables after a suitable Lagrange multiplier is derived by Eq.~(\ref{4.14}).
The above action can be recast as
\begin{equation}\label{10}
S=4\pi\int dt \left\{ a^3f(R)-\lambda\left[R+6\left(\frac{\ddot{a}}{a}+\frac{\dot{a}^2}{a^2}+\frac{k}{a^2}\right)\right]\right\}\,{,}
\end{equation}
where the Lagrange multiplier $\lambda$ can be obtained by varying the above action with respect to $R$, giving
\begin{equation}\label{11}
\lambda=a^3f'(R)\,.
\end{equation}
 After an integration by parts, the Lagrangian $\mathcal{L}$ assumes the form 
\beq
\mathcal{L}=a^3\left[f(R)-Rf'(R)\right]+6a^2f''(R)\dot{a}\dot{R}+6f'(R)a\dot{a}^2-6kaf'(R),            \label{4.4}
\eneq
from which it is straightforward to obtain the Euler-Lagrange equations 
\beq
\frac{d}{dt}\frac{\partial\mathcal{L}}{\partial\dot{a}}=\frac{\partial\mathcal{L}}{\partial a}\,,             \label{4.9} 
\eneq 
\beq
\frac{d}{dt}\frac{\partial\mathcal{L}}{\partial\dot{R}}=\frac{\partial\mathcal{L}}{\partial R}\,,             \label{4.22}     
\eneq
which, with some algebra, correspond to the cosmological equations~\cite{Capozziello:2014hia} 
\begin{equation}
\label{13}
2\left(\frac{\ddot{a}}{a}\right)+\left(\frac{\dot{a}}{a}\right)^2+\frac{k}{a^2}=-p_{(tot)},
\end{equation}
and
\begin{equation}
\label{14}
f''(R)\left\{R+6\left[\frac{\ddot{a}}{a}+\left(\frac{\dot{a}}{a}\right)^2+\frac{k}{a^2}\right]\right\}=0\,.
\end{equation}
In particular, Eq.~(\ref{14}) can be  interpreted in terms of the Lagrange multiplier definition, guaranteeing the consistency of the approach.
Furthermore, the Hamiltonian constraint 
\beq
E_\mathcal{L}=\frac{\partial\mathcal{L}}{\partial\dot{a}}\dot{a}+\frac{\partial\mathcal{L}}{\partial\dot{R}}\dot{R}-\mathcal{L}=0\,,       \label{4.37}
\eneq
gives
\begin{equation}
\label{15}
\left(\frac{\dot{a}}{a}\right)^2+\frac{k}{a^2}=\frac{1}{3}\rho_{(tot)}\,.
\end{equation}
Here,  for the sake of completeness, we added also standard matter that, in the Jordan frame, is minimally coupled to the geometry. By writing
\begin{equation}
\label{17}
p_{(tot)}=p_{(curv)}+p_{(m)}\,, \,\,\,\,\, \rho_{(tot)}=\rho_{(curv)}+\rho_{(m)}\,,
\end{equation}
we put in evidence both curvature and matter contributions, where we have inserted the non-minimal coupling factor $1/f'(R)$ into the definition of the matter terms.  From $T^{(curv)}_{\mu\nu}$, it is easy to get the {\it curvature pressure} 
\begin{equation}
\label{18}
p_{(curv)}=\frac{1}{f'(R)}\left\{2\left(\frac{\dot{a}}{a}\right)\dot{R}f''(R)+\ddot{R}f''(R)+\dot{R}^2f'''(R)
-\frac{1}{2}\left[f(R)-Rf'(R)\right] \right\}\,,
\end{equation}
and the {\it  curvature density}
\begin{equation}
\label{19}
\rho_{(curv)}=\frac{1}{f'(R)}\left\{\frac{1}{2}\left[f(R)-Rf'(R)\right]
-3\left(\frac{\dot{a}}{a}\right)\dot{R}f''(R) \right\}\,.
\end{equation}
Exact solutions for this dynamical  system  can be achieved by searching for Noether symmetries (see the Appendix \ref{six} for a detailed description of the method)~\cite{Capozziello:2008ch}.  Here we want to show that such symmetries are related to the 
scale factor duality for the $f(R)$ Lagrangians discussed in the above Section.  In other words, there exist models in $f(R)$ gravity with the same feature of the string-dilaton cosmology at the level of the Lagrangian. The presence of Noether symmetries determines this result.  In our further considerations, we will discard the standard matter contribution that is not necessary to the aims of this paper. The contribution of  perfect fluid matter to Noether symmetries in $f(R)$ gravity is discussed in Ref.~\cite{Capozziello:2008ch}. 

\subsection {Invariant $f(R)$ models for $a\rightarrow 1/a$ transformations}

Let us use the constant of motion (\ref{4.34}), obtained from the Noether Symmetry Approach, in order to select a class of Lagrangians where scale factor duality is present. 
For simplicity reasons let us set $c_2=\bar{c}=0$, then Eqs.~(\ref{4.35}) and~(\ref{4.36}) become
\beq
\alpha=c_1a\,,                                                        \label{5.0}
\eneq  
\beq
\beta=-3c_1\frac{f'(R)}{f''(R)}\,.                                    \label{5.1}
\eneq
Inserting these definitions into the constant of motion (\ref{4.34}) and choosing $\Sigma_0=0$, we obtain:
\beq
6c_1f''(R)a^3\dot{R}+12c_1f'(R)a^2\dot{a}-18c_1f'(R)a^2\dot{a}=0\,.
\eneq
Immediately  we  get
\beq
f''(R)\dot{R}=f'(R)\left(\frac{\dot{a}}{a}\right).                                \label{5.2}
\eneq
In order to select $f(R)$ models which undergo scale factor duality transformations, we insert Eq.~(\ref{5.2}) into the point-like Lagrangian (\ref{4.4}). In the following we choose $k=0$ (spatially flat metric) for the sake of simplicity\footnote{Noether symmetries exist also in the cases $k\neq 0$ as it is shown in Ref.~\cite{Capozziello:2008ch}.}. We obtain the simpler Lagrangian
\beq
\mathcal{L}=a^3\left[f(R)-f'(R)R\right]+12a\dot{a}^2f'(R).           \label{5.3}
\eneq  
Keeping in mind Eq.~(\ref{3.23}), the previous Lagrangian $\mathcal{L}$ can be rewritten as follows:
\begin{equation} 
\mathcal{L}=a^3\left[(2\Lambda e^{2\phi}-4\Lambda R e^{2\phi}{\phi}'(R))+48\left(\frac{\dot{a}}{a}\right)^{2} \Lambda e^{2\phi}\phi'(R) \right].
\label{resco}   
\end{equation}
Notice that, with abuse of notation, we are indicating the Ricci scalar $\tilde{R}$ built with the metric $\tilde{g}_{\mu\nu}$, simply as $R$.
 
Being  $a^{3}=e^{3\ln(a)}$, we can rewrite Eq.~(\ref{resco}) as
 \begin{equation} 
\mathcal{L}=\left[(2\Lambda e^{2\left(\phi +\frac{3}{2}\ln(a)\right)}-4\Lambda R e^{2\left(\phi +\frac{3}{2}\ln(a)\right)}{\phi}'(R))+48\left(\frac{\dot{a}}{a}\right)^{2} \Lambda e^{2\left(\phi +\frac{3}{2}\ln(a)\right)}\phi'(R) \right]\,. 
\label{resco2}
\end{equation}
Applying the dilaton shift as similarly done in Eqs.~(\ref{conf}) and~(\ref{cosmdual}), we have  
\begin{equation}
\phi \rightarrow {\tilde\phi}=\phi + \frac{3}{2}\ln (a)\,,
\label{sco}
\end{equation}
and the Lagrangian becomes
\begin{equation} 
\mathcal{L}=\left[2\Lambda e^{2{\tilde{\phi}}}-4\Lambda R e^{2\tilde{\phi}}\tilde{\phi}'(R)\right]+48\left(\frac{\dot{a}}{a}\right)^{2} \Lambda e^{2\tilde{\phi}}\tilde{\phi}'(R)\,, 
\label{resco3}
\end{equation}
which can be recast as 
\beq
\mathcal{L}=\left[f(R)-f'(R)R\right]+12\left(\frac{\dot{a}}{a}\right)^2f'(R)\,.        \label{5.4}
\eneq
The latter is invariant under the scale factor duality $a\rightarrow 1/a$ and, furthermore, gives rise to the same dynamics as the Lagrangian in Eq.~(\ref{5.3}). After these considerations, we point out that the presence of the Noether symmetry gives rise to the scale factor duality invariance.
 
\subsection {Cosmological solutions}

Exact cosmological solutions can be obtained starting from the previous results. Considering the Ricci scalar curvature in Eq.~(\ref{4.14}) for a spatially flat metric, we can compare it with the equation for $R$ which can be obtained using the Lagrangian in Eq.~(\ref{5.4}), that is:
\beq
R=12\left(\frac{\dot{a}}{a}\right)^2,                                        \label{5.17}           
\eneq
which is clearly invariant under scale factor duality. 
We get
\beq
a\ddot{a}+3\dot{a}^2=0\,,                                            \label{5.18}
\eneq
obtaining  the solution 
\beq
a(t)=\sqrt[4]{c_1t+c_2}\,,                     \label{5.19}
\eneq
where $c_1$ and $c_2$ are integration constants. 
   
The dynamical equation for $a(t)$ is obtained from the Lagrangian (\ref{5.4}) as
\beq
a\ddot{a}f'(R)+a\dot{a}\dot{R}f''(R)-\dot{a}^2f'(R)=0\,,
\eneq 
which, by using Eqs.~(\ref{5.17}) and~(\ref{5.19}), can be recast in the form
\beq
f'(R) + 2 R f''(R) = 0\,. \label{eqscale}
\eneq
Moreover, the Hamiltonian constraint in Eq.~(\ref{4.37}) gives
\begin{equation}
f(R)-2Rf'(R)=0\,.
\label{hamconst}
\end{equation}
Finally, by using Eqs.~(\ref{eqscale}) and~(\ref{hamconst}), the functional form of $f(R)$ is fixed to be 
\beq
f(R)=f_0 R^{\frac{1}{2}},
\eneq 
where $f_0$ is an integration constant. 

\subsection{General duality transformations as reflections}
 
It is possible, by using the Noether Symmetry Approach, to obtain large classes of theories where scale factor duality can be represented as  a reflection~\cite{Capozziello:1993vs,Capozziello:1993vr,Capozziello:1993ts}. In the specific case of $f(R)$ gravity, let us consider again the Lagrangian (\ref{5.4}) and perform the following redefinitions:
\beq
R = Ae^{-\phi}, \,\,\,\,\,   f(R) = e^{-2\phi}F(\phi),   \label{change}
\eneq
where $\phi$ is the dilaton field and $A$ is an arbitrary constant. 
Then we obtain
\beq
f'(R)=\frac{df}{dR}=\frac{df}{d\phi}\frac{d\phi}{dR}=-A^{-1}e^{-\phi}\left[F'(\phi)-2F(\phi)\right],    
\eneq
where $F'(\phi)=dF(\phi)/d\phi$.
Inserting these transformations into the Lagrangian in Eq.~(\ref{5.4}), the latter becomes
\beq
\mathcal{L}=\left[e^{-2\phi}\left(F'(\phi)-F(\phi)\right)\right]-12A^{-1}e^{-\phi}\left(\frac{\dot{a}}{a}\right)^2\left[F'(\phi)-2F(\phi)\right].               \label{5.27}
\eneq  
We perform now the following transformation:
\beq
\phi=g(w),             \label{5.28}
\eneq
where $g(w)$ is a general function of $w$~\cite{Capozziello:1993vr,Capozziello:1993ts}.      
So, the Lagrangian (\ref{5.27}) can be rewritten as
\beq
\mathcal{L}= e^{-2g(w)}\left[F'(\phi)-F(\phi)\right]-12A^{-1}e^{-g(w)}\left(\frac{\dot{a}}{a}\right)^2\left[F'(\phi)-2F(\phi)\right].   \label{5.29}
\eneq
Assuming that $[F'(\phi)-F(\phi)]=\gamma$, where $\gamma$ is a constant, we determine a possible form of the function $F(\phi)$ and then of $f(R)$:
\beq
F'(\phi)-F(\phi)=\gamma \,\, \Rightarrow \,\, F(\phi)=\left(\xi e^{\phi}-\gamma\right),   \label{change2}
\eneq
where $\xi$ is an integration constant. Then, from  Eqs.~(\ref{change}), we get the correspondent functional form of $f(R)$ which is
\beq
f(R)=\frac{1}{A}\left(\xi R-\frac{\gamma}{A} R^2\right).
\eneq
So, the Lagrangian  (\ref{5.29}) becomes 
\beq
\mathcal{L}= \gamma e^{-2g(w)}-12 A^{-1} \left(\frac{\dot{a}}{a}\right)^2\left(2\gamma e^{-g(w)} - \xi \right).  \label{5.31}
\eneq
Let us complete the change of variables $\left(a,\phi\right)\rightarrow\left(z,w\right)$ by using the following transformation:
\beq
z=\ln\frac{a}{\sigma}-\mu(w),
\eneq
where $\mu(w)$ is a generic function and $\sigma$ is a non-zero positive parameter. Immediately we have 
\beq
a=\sigma e^ze^{\mu(w)},  \hspace{1cm}     \dot{a}=\sigma e^ze^{\mu(w)}\left(\dot{z}+\mu_w \dot{w}\right).  
\eneq
Then, the Lagrangian  (\ref{5.31}) takes the form
\beq
\mathcal{L}= \gamma e^{-2g(w)}-12 A^{-1} \left(\dot{z}+\mu_w \dot{w}\right)^2\left(2\gamma e^{-g(w)} - \xi \right).   \label{5.32}
\eneq 
Let us stress  that the functions $g(w)$ and $\mu(w)$ are completely generic. If we assume that $g(w)$ and $\mu(w)$ are, respectively, even and odd functions, that is 
\beq
g(-w)=g(w), \,\,\,\,  \mu(-w)=-\mu(w),                       \label{5.33}
\eneq
then the Lagrangian (\ref{5.32}) is form invariant under the change 
\beq
w\rightarrow-w, \,\,\,\, z\rightarrow-z,
\eneq
these transformations being equivalent to $a\rightarrow 1/a$, if additionally $\sigma \rightarrow 1/\sigma$. We then conclude that, in the $\{z,w\}$ space, the duality invariance  is a reflection, i.e. an invariance under parity transformation in the considered configurations space. Furthermore the Lagrangian (\ref{5.32}) represents a class of models, being the two functions $g(w)$ and $\mu(w)$ completely arbitrary. Finally the Lagrangian (\ref{5.32}) is cyclic in $z$ so that, in the $\{z,w\}$ space, the dynamics is reduced  and a constant of motion exists. 
In other words, the Noether Symmetry gives rise to a reflection transformation that generalizes scale factor duality.

\section{Conclusions and perspectives}
\label{five}

Modified theories of gravity are a useful paradigm to cure shortcomings of GR at ultraviolet and infrared scales, due to the lack of a full Quantum Gravity theory.
Despite the success of addressing a lot of phenomenology, ranging from inflation~\cite{Starobinsky:1980te}, accelerated behavior of present universe~\cite{Capozziello:2007ec,Capozziello:2011et}, up to to self-gravitating structures~\cite{Capozziello:2012ie}, they should be framed into some fundamental theory in order to achieve a self-consistent picture of gravity at all scales. 

Due to this state of art, a sort of {\it reconstruction technique} is extremely useful. The philosophy is to find out some physically relevant features that allow to relate ETGs to some fundamental theory as Superstring, Supergravity et al. 

In this paper, we have considered the straightforward extension of GR, that is $f(R)$ gravity, where the assumption $f(R)=R$ of the Hilbert-Einstein action is relaxed. 
As it is well known, such an extended theory has revealed to be very useful in the last decade in order to address a lot of open problems in astrophysics and cosmology~\cite{Capozziello:2007ec}. Here, we have investigated the possibility that scale factor duality, one of the main characteristics of string-dilaton theory, holds also for some class of $f(R)$ models.

Recasting the tree-level dilaton-graviton effective action of bosonic String Theory into the $f(R)$ theory, we have seen that the dilaton field $\phi$ has a geometrical interpretation in terms of the Ricci scalar curvature. The specific form of the $f(R)$ function is given by the dilaton non-minimal coupling and the dilaton potential by a conformal transformation. 
Moreover, the possibility to achieve scale factor duality clearly depends on the presence of Noether symmetry in the $f(R)$ Lagrangian. 

This result is more general than the case of the tree-level effective bosonic String Theory, since the same scale factor duality (a discrete transformation) can bring back to a Noether symmetry (a continuous transformation).
Furthermore, the Noether symmetry allows to reduce the dynamics, to obtain Lagrangians which are invariant under duality transformations and then to achieve exact cosmological solutions.

The issue taken into account in this paper is more general since the same program could be applied to other effective theories and considering other fundamental features in addition to the duality. This will be the argument of further investigations. 

\section*{Acknowledgements}
SC acknowledges INFN Sez. di Napoli ({\it Iniziativa Specifica} QGSKY) for financial support. The research of DV leading to these results has received funding from the European Research Council under the European Community's Seventh Framework Programme (FP7/2007-2013, Grant Agreement No. 307934). The Authors wish to thank Stefano Bellucci and Mariafelicia De Laurentis for useful discussions and suggestions on the topic. 

\appendix

\section{The Noether Symmetry Approach} 
\label{six}

Dynamics given by (\ref{4.4}) can be reduced and solved searching for  Noether symmetries. In principle, the Noether Symmetry  Approach allows to select dynamical  models where conserved charges come out allowing the reduction~\cite{Capozziello:1996bi, Capozziello:1999xs,Capozziello:2007id,Capozziello:2007wc,Capozziello:2008ch,Capozziello:2012iea}. 
The approach can be sketched in the following way.
 
Let us consider a canonical, non-degenerate point-like Lagrangian $\mathcal{L}(q^i,\dot{q}^i)$ that is
\beq
\frac{\partial\mathcal{L}}{\partial\lambda}=0, \hspace{1.5cm}  \mbox{det} H_{ij}\equiv \mbox{det} \left\|\frac{\partial^2\mathcal{L}}{\partial \dot{q}^i\partial \dot{q}^j}\right\|\neq0,
\eneq
where $H_{ij}$ is the Hessian matrix. The  dot is the derivative with respect to the affine parameter $\lambda$ (in our case  to the cosmic time). In general,  $\mathcal{L}$ has  the form 
\beq
\mathcal{L}=T(\textbf{q},\dot{\textbf{q}})-V(\textbf{q}),                   \label{4.39} 
\eneq
where \textit{T} and \textit{V} are, respectively, the kinetic and potential terms. The energy function associated to  $\mathcal{L}$ is
\beq
E_\mathcal{L}\equiv\frac{\partial\mathcal{L}}{\partial \dot{q}^i}\dot{q}^i-\mathcal{L}\,,
\eneq  
which is a constant of motion. Since cosmological problems have a finite number of degrees of freedom, one can take into account  {\it point transformations}. Invertible coordinate transformations  $Q^i=Q^i(\textbf{q})$ induce transformations of the velocities, that is 
\beq
\dot{Q}^i(\textbf{q})=\frac{\partial Q^i}{\partial q^j}\dot{q}^j\,,   \label{4.23} 
\eneq
and the Jacobian of the transformation $\mathcal{J}= \mbox{det} \left\|\partial Q^i/\partial q^j\right\|$ is assumed to be non-zero. 

In general, an infinitesimal point transformation is represented by a vector field 
\beq
\textbf{X}=\alpha^i(\textbf{q})\frac{\partial}{\partial q^i}+\left(\frac{d}{d\lambda}\alpha^i(\textbf{q})\right)\frac{\partial}{\partial \dot{q}^i}.   \label{4.24}       
\eneq
A function $F(\textbf{q}, \dot{\textbf{q}})$ is invariant under the transformation \textbf{X} if
\beq
L_X F\equiv\alpha^i(\textbf{q})\frac{\partial F}{\partial q^i}+\left(\frac{d}{d\lambda}\alpha^i(\textbf{q})\right)\frac{\partial}{\partial \dot{q}^i}F=0,
\eneq
where $L_XF$ is the Lie derivative of \textit{F}.   
In particular, the condition  $L_X\mathcal{L}=0$ means that the vector \textbf{X} is a symmetry for the Lagrangian $\mathcal{L}$.
Let us consider now a Lagrangian $\mathcal{L}$ and the related  Euler-Lagrange equations
\beq
\frac{d}{d\lambda}\frac{\partial\mathcal{L}}{\partial \dot{q}^j}-\frac{\partial\mathcal{L}}{\partial q^j}=0.                \label{4.25}
\eneq
Considering the vector \textbf{X} in Eq.~(\ref{4.24}) and contracting Eq.~(\ref{4.25}) with $\alpha^j$'s, gives
\beq
\alpha^j\left(\frac{d}{d\lambda}\frac{\partial\mathcal{L}}{\partial \dot{q}^j}-\frac{\partial\mathcal{L}}{\partial q^j}\right)=0.  \label{4.26}\eneq
Since 
\beq
\alpha^j\frac{d}{d\lambda}\frac{\partial\mathcal{L}}{\partial \dot{q}^j}=\frac{d}{d\lambda}\left(\alpha^j\frac{\partial\mathcal{L}}{\partial \dot{q}^j}\right)-\left(\frac{d\alpha^j}{d\lambda}\right)\frac{\partial\mathcal{L}}{\partial \dot{q}^j}\,,
\eneq
from Eq.~(\ref{4.26}), it follows
\beq
\frac{d}{d\lambda}\left(\alpha^j\frac{\partial\mathcal{L}}{\partial \dot{q}^j}\right)=L_X\mathcal{L}.
\eneq
The  conseguence is the \textit{Noether theorem}:

If $L_X\mathcal{L}=0$, then the function
\beq
\Sigma_0=\alpha^k\frac{\partial\mathcal{L}}{\partial \dot{q}^k},                                                \label{4.27} 
\eneq
is a constant of motion.

Let us now consider the Lagrangian in Eq.~(\ref{4.39}). Since \textbf{X} is of the form (\ref{4.24}), $L_X\mathcal{L}$ is a homogeneous polynomial of second degree in the velocities plus an inhomogeneous term in the $q^i$. This polynomial has to be identically zero and  each coefficient must be independently zero. If $n$ is the dimension of the configuration space, we obtain $\left\{1+n(n+1)/2\right\}$ partial differential equations. This system is overdetermined, therefore the Noether Symmetry Approach can be used to select the functions which assign the models. In the case of modified gravity theories, these functions are couplings and potentials~\cite{Capozziello:1996bi}.
 
Considering the specific case which we are discussing,  $f(R)$ cosmology, the configuration space is $\mathcal{Q}=\left\{a,R\right\}$, and  the tangent space  is $\mathcal{TQ}=\left\{a,\dot{a},R,\dot{R} \right\}$.
 The Lagrangian is an application
\beq
\mathcal{L}:\mathcal{TQ}\rightarrow\mathcal{R},
\eneq
where $\mathcal{R}$ are the  real numbers.
The generator of symmetry is
\beq
\textbf{X}=\alpha\frac{\partial}{\partial a}+\beta\frac{\partial}{\partial R}+\dot{\alpha}\frac{\partial}{\partial \dot{a}}+\dot{\beta}\frac{\partial}{\partial \dot{R}}.           \label{4.28}
\eneq
As discussed above, a symmetry exists if the equation $L_X\mathcal{L}=0$ has solutions.  In other words, a symmetry exists if at least one of the functions $\alpha$ or $\beta$ in Eq.~(\ref{4.28}) is different from zero.
Going to our specific case, the Lagrangian (\ref{4.4}),  and setting to zero the coefficients of the terms $\dot{a}^2$, $\dot{R}^2$, $\dot{a}\dot{R}$, we obtain the following system of equations,  linear in $\alpha$ and $\beta$,
\begin{eqnarray}
&&f'(R)\left(\alpha+2a\partial_a \alpha\right)+af''(R)\left(\beta +a\partial_a\beta\right)=0,         \label{4.29}\\
&&a^2f''(R)\partial_R\alpha=0,                                                                        \label{4.30}\\ 
&&2f'(R)\partial_R\alpha+f''(R)\left(2\alpha+a\partial_a\alpha+a\partial_R\beta\right)+a\beta f'''(R)=0,            \label{4.31}     
\end{eqnarray}
and, finally, setting to zero the remnant terms, we obtain the constraint
\beq
3\alpha\left(f(R)-Rf'(R)\right)-a\beta Rf''(R)-\frac{6k}{a^2}\left(\alpha f'(R)+a\beta f''(R)\right)=0.\label{pippo}
\eneq
Solutions of Eqs.~(\ref{4.29})-(\ref{pippo}) exist if explicit forms of $\alpha$ and $\beta$ are found. In other words,  if at least one of the functions $\alpha$ and $\beta$ are  different from zero, a Noether symmetry exists. 
If $f''(R)\neq 0$, Eq.~(\ref{4.30}) can be immediately solved:
\beq
\alpha=\alpha(a).
\eneq
We do not take into account the case $f''(R)=0$ because it corresponds to standard GR. We can rewrite Eqs.~(\ref{4.29}) and~(\ref{4.31}) as follows:
\beq
f'(R)\left(\alpha+2a\frac{d\alpha}{da}\right)+af''(R)\left(\beta+a\partial_a\beta\right)=0,                     \label{4.32}
\eneq  
\beq
f''(R)\left(2\alpha+a\frac{d\alpha}{da}+a\partial_R\beta\right)+a\beta f'''(R)=0.                               \label{4.33} 
\eneq
Since $f=f(R)$, then $\partial f/\partial a=0$; then it is possible to solve Eq.~(\ref{4.33}) by writing it as:
\beq
\partial_R(\beta f''(R))=-f''(R)\left(2\frac{\alpha}{a}+\frac{d\alpha}{da}\right),
\eneq
and, by integration,  its general solution is
\beq
\beta=-\left[\frac{2\alpha}{a}+\frac{d\alpha}{da}\right]\frac{f'(R)}{f''(R)}+\frac{h(a)}{f''(R)}.
\eneq
Therefore we find that Eq.~(\ref{4.32}) gives
\beq
f'(R)\left[\alpha-a^2\frac{d^2\alpha}{da^2}-a\frac{d\alpha}{da}\right]+a\left[h+a\frac{dh}{da}\right]=0,
\eneq 
which has the solution
\beq
\alpha=c_1a+\frac{c_2}{a}  \hspace{1cm}   \mbox{and}     \hspace{1cm}     h=\frac{\bar{c}}{a},           \label{4.35}
\eneq
where, $a$ being dimensionless, $c_1$ and $c_2$ have the same dimensions. We can also fix $\alpha$ to be dimensionless and this fixes the dimensions of $\beta$ to be $[\beta]=M^2$. Then also $[\bar{c}]=M^2$, so finally we have:
\beq
\beta=-\left[3c_1+\frac{c_2}{a^2}\right]\frac{f'(R)}{f''(R)}+\frac{\bar{c}}{af''(R)}.                     \label{4.36}
\eneq
This Noether symmetry implies the existence of a constant of motion. From Eq.~(\ref{4.27}) and the Lagrangian~(\ref{4.4}) we obtain:
\beq
\alpha\left(6f''(R)a^2\dot{R}+12f'(R)a\dot{a}\right)+\beta\left(6f''(R)a^2\dot{a}\right)=\Sigma_0\,,         \label{4.34}
\eneq
that we have used in the above considerations on duality.


\end{document}